\documentclass[twocolumn,pacs,aps,prl,floatfix,superscriptaddress]{revtex4-1}
\usepackage{amsmath,amssymb}
\usepackage{graphicx,float}
\usepackage{times,txfonts}
\usepackage{color}
\usepackage{multirow}
\usepackage{bbold}
\usepackage{color}
\usepackage{soul}
\usepackage{hyperref}
\usepackage{times,txfonts}
\usepackage{natbib}

\newcommand{\abs}[1]{\left| #1 \right|}

\newcommand{\ket}[1]{\left| #1 \right\rangle}
\newcommand{\braket}[2]{\left\langle {#1{\left| \vphantom{#1 #2} \right.} #2} \right\rangle}
\newcommand{\qo}[1]{``#1''}

\renewcommand{\epsilon}{\varepsilon}
\renewcommand{\phi}{\varphi}

\usepackage{sistyle}
\usepackage{soul}
\definecolor{lightblue}{RGB}{185,210,248}
\sethlcolor{lightblue}

\begin{document}
%
\title{Real-time imaging of spin-to-orbital angular momentum quantum state teleportation}
\author{Manuel Erhard}
\altaffiliation [Also at ]{Faculty of Physics, University of Vienna, Boltzmanngasse 5, A-1090 Vienna, Austria}
\affiliation{Department of Physics, University of Ottawa, 150 Louis Pasteur, Ottawa, Ontario, K1N 6N5 Canada}
\author{Hammam Qassim}
\affiliation{Department of Physics, University of Ottawa, 150 Louis Pasteur, Ottawa, Ontario, K1N 6N5 Canada}
\author{Harjaspreet Mand}
\affiliation{Department of Physics, University of Ottawa, 150 Louis Pasteur, Ottawa, Ontario, K1N 6N5 Canada}
\author{Ebrahim Karimi}
\email{ekarimi@uottawa.ca}
\affiliation{Department of Physics, University of Ottawa, 150 Louis Pasteur, Ottawa, Ontario, K1N 6N5 Canada}
\author{Robert W Boyd}
\affiliation{Department of Physics, University of Ottawa, 150 Louis Pasteur, Ottawa, Ontario, K1N 6N5 Canada}
\affiliation{Institute of Optics, University of Rochester, Rochester, New York, 14627, USA}

\begin{abstract}
Quantum teleportation is a process in which an unknown quantum state is transferred between two spatially separated subspaces of a bipartite quantum system which share an entangled state and communicate classically. In the case of photonic states, this process is probabilistic due to the impossibility of performing a two-particle complete Bell state analysis with linear optics. In order to achieve a deterministic teleportation scheme, harnessing other degrees of freedom of a single particle, rather than a third particle, has been proposed. Indeed, this leads to a novel type of deterministic teleportation scheme, the so-called hybrid teleportation. Here we report the first realization of photonic hybrid quantum teleportation from spin-to-orbital angular momentum degrees of freedom. In our scheme, the polarization state of photon \emph{A} is transferred to orbital angular momentum of photon \emph{B}. The teleported states are visualized in  real-time by means of an intensified CCD camera. The quality of teleported states is verified by performing quantum state tomography, which confirms an average fidelity higher than 99.4\%. We believe this experiment paves the route towards a novel way of quantum communication in which encryption and decryption are carried out in naturally different Hilbert spaces, and therefore may provide means of enhancing security. 
\end{abstract}
\maketitle

Entanglement is one of the most interesting aspects of quantum mechanics and is at the heart of several quantum paradoxes, such as the Einstein-Podolsky-Rosen (EPR) paradox~\cite{epr}, Hardy paradox~\cite{hardy:93}, and Leggett's inequalities~\cite{leggett:03}. One of the features that is made possible via entanglement is the ability to teleport arbitrary quantum states. In general, a quantum teleportation scheme describes how to transmit a quantum state between two spatially separated participants, usually called Alice and Bob -- hereafter referred to as \emph{A} and \emph{B}, respectively. In 1997, Bennett and coworkers proposed the first quantum state teleportation scheme, which was based on three spin-half particles~\cite{tele1}. In their scheme, particles one and two are in an EPR-entangled state in the spin degree of freedom (DOF). Particle three, Charlie (\emph{C}), is in a quantum state $\ket{\Phi}$, which in general is unknown. \emph{A} has particles one and three while \emph{B} has particle two of the entangled pair. In order to transmit the quantum state $\ket{\Phi}$ from \emph{A} to \emph{B}, \emph{A} performs a joint Bell-state measurement on particles one and three. She then sends \emph{B} a classical message with the measurement outcome, which contains two classical bits. With this information \emph{B} is able to perform a unitary operation to reconstruct the quantum state $\ket{\Phi}$. This scheme needs three particles and uses their spin DOF to transmit the non-classical information. Later on, however, different theoretical and experimental teleportation schemes based on use of multiple particles and different degrees of freedom were proposed in the literature~\cite{zeilinger97,demartini98,barreiro:10,khoury1,krauter2013}. Among those, schemes that are based on different degrees of freedom, i.e. hybrid teleportation, received particular attention. This is due to the deterministic nature of these schemes, in contrast to implementations based on multiple particles -- recall the impossibility of performing two-particle complete Bell-state measurements with linear optics. In addition, combining different degrees of freedom of a single particle has recently received immense attention, as it provides a novel way to perform high dimensional quantum key distribution~\cite{barreiro:05}, superdense coding~\cite{barreiro:08,milione:13} and quantum metrology~\cite{vincenzo:13}. Spin and orbital angular momentum (OAM) of light, associated respectively with the vectorial nature and helical phase-fronts of optical beams, have been thoroughly examined, alongside the very recently investigated radial index of Laguerre-Gauss modes~\cite{karimi:14}. However, it is worth mentioning that an experimental realization of hybrid teleportation remains unexploited.
\begin{figure}[h]
\includegraphics[scale=0.25]{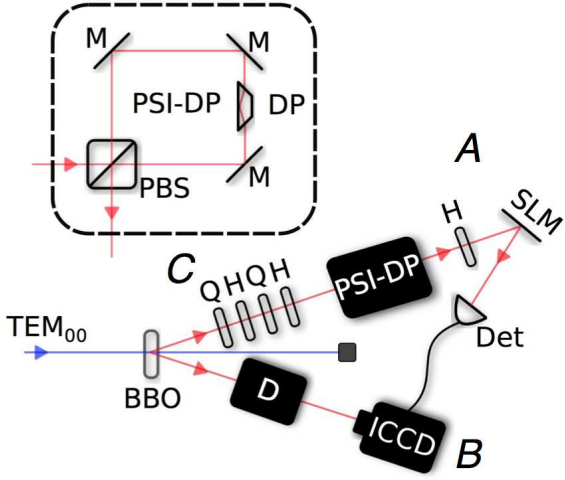}
	\caption{\label{fig:fig1} (Online color) A quasi-cw $150$ mW UV laser with a repetition rate of $100$~MHz at $355$~nm operating at the fundamental $\text{TEM}_{00}$ mode pumps a $\beta$-barium borate (BBO) crystal cut for type-I phase matching. This generates photon pairs that are entangled in their OAM degree of freedom. A sequence of quarter- (QWPs) and half-wave plates (HWPs) are used to prepare the polarization of photon \emph{A} in an arbitrary state. The polarizing Sagnac interferometer with a Dove Prism (PSI-DP), shown in upper inset, in combination with the HWP, spatial light modulator (SLM) and single mode optical fibre perform the Bell-state measurement. Our SLM, HoloEye Pluto, is polarization sensitive and diffracts only horizontally polarized photons, thus acting as a polarizer. In fact, an SLM with a proper hologram and a single mode optical fiber post-select the OAM Hilbert sub-space, where only photons with a flattened wave-front, conjugate of the hologram displayed on SLM, can be coupled into the fiber and detected by an avalanche photon detector (APD). The signal from the detector triggers the ICCD camera to record the spatial distribution of photon \emph{B}. In order to reduce the noise, photons are filtered with $10$ nm bandpass filters before the detector and ICCD camera. During the preparation and measurement of photon \emph{A}, photon \emph{B} circulates in the delay line D, which is appropriately designed to compensate for the electronic delay. In order to measure the fidelity of the teleported states, the ICCD camera is replaced by another SLM and the coincidence counts between \emph{A}'s and \emph{B}'s detectors are measured.}
\end{figure}

In this Letter, we show experimentally how to teleport a polarization state of one photon to a different Hilbert space, here the OAM, of its entangled partner photon.  Additionally, an unknown quantum state can be teleported from polarization to the OAM degree of freedom. Hence, in contrast to previous teleportation schemes where a third \emph{ancilla} photon carries the unknown state, the present one involves two photons entangled in the OAM DOF, with the unknown state being carried in the polarization DOF of one of the photons. The presented teleportation protocol is potentially useful in quantum computing, quantum cryptography and quantum networks. It makes it possible to connect different physical systems like photon-based quantum computers with photonic quantum memories, for example~\cite{nicolas2014quantum}. It also shows experimentally that quantum information in general can be transmitted between completely different physical properties of spatially dislocated particles, as long as they are entangled. 

Let us now briefly discuss our implemented teleportation scheme, which is a merger of two different proposals reported in Ref.~\cite{khoury1,chen:09}. In our scheme, \emph{A} and \emph{B} share a pair of photons entangled in their OAM degree of freedom. \emph{C} sets the polarization of \emph{A}'s photon to an arbitrary state, which is unknown to \emph{A} as well. \emph{A} then performs a full Bell-state measurement on the spin and OAM sub-space of her photon, and transfers the outcome results that is two classical bits to \emph{B} via a classical channel. As we will show, this leaves \emph{B}'s photon in a superposition of orthogonal OAM states, which is determined by the polarization, state and the value of OAM defined by the Bell-state measurement. According to the classically communicated measurement outcome, \emph{B}  applies a unitary operator from the set $\{\hat{\mathbb{1}},\hat{\sigma}_x,i\hat{\sigma}_y,\hat{\sigma}_z\}$, which are the Pauli matrices for a bi-dimensional system. During the preparation, the measurement process of photon \emph{A} and the transmission of the classical bits to \emph{B}, photon \emph{B} circulates in a delay line. To measure the OAM superposition of photon \emph{B} an intensified charge coupled device (ICCD) camera is used at the end of the delay line to record its transverse intensity distribution.  

Photon pairs entangled in position and anti-correlated in momentum space (EPR states) are generated via a spontaneous parametric downconversion (SPDC) process in a $\beta$-barium borate (BBO) crystal cut for type-I collinear phase matching. Under satisfying the phase matching condition and assuming a pump beam with a Gaussian profile, the angular momentum is conserved and the state can be written as~\cite{conservation-oam-in-spdc}:
\begin{equation}  \label{SPDC_state}
    \ket{\chi}=\sum_{\ell=0}^{\infty}c_{\ell}\Bigg(\ket{-\ell}_A\ket{+\ell}_B+\ket{+\ell}_A\ket{-\ell}_B\Bigg)\otimes\ket{H}_A\ket{H}_B,
\end{equation}
where $c_\ell$ is a constant depending on crystal and pump properties, and $\ket{\ell}_i$ represents the OAM state of photon $i$, which is $\ell$ in the direction of propagation in units of $\hbar$ -- the reduced Planck's constant~\cite{mair:01}. $\ket{H}$ refers to horizontal polarization of photon \emph{A} or \emph{B}. Subsequently \emph{C} prepares an arbitrary state in the polarization DOF of photon \emph{A} to be teleported to \emph{B}. This is achieved with a series of half- and quarter-wave plates  (HWP) \& (QWP) shown in Fig.~\ref{fig:fig1}, which does an arbitrary SU(2) transformation on the polarization state of photon \emph{A}. Recall that the polarization state of photon \emph{A} and \emph{B} is in a product state as shown in Eq.~(\ref{SPDC_state}). The polarization transformation of photon \emph{A} is then given by $\ket{H}_A\rightarrow \alpha\ket{H}_A+\beta\ket{V}_A$ where $\alpha,\beta$ are arbitrary complex numbers with $|\alpha|^2+|\beta|^2=1$. $\ket{H}$ and $\ket{V}$ refer to horizontal and vertical polarizations, respectively. To see the action of a Bell measurement on the SPDC state $\ket{\chi}$ of Eq.~(\ref{SPDC_state}), we define the four Bell states for a single photon, but in two bi-dimensional Hilbert spaces of polarization $\{\ket{H},\ket{V}\}$ and OAM subspace of $\{\ket{+\ell},\ket{-\ell}\}$ :
\begin{eqnarray}\label{bell_states}
	\ket{\Phi^{\pm}_\ell}_A&=&\frac{1}{\sqrt{2}}\Big(\ket{h_\ell,H}_A\pm\ket{v_\ell,V}_A\Big)\nonumber\\
	\ket{\Psi^{\pm}_\ell}_A&=&\frac{1}{\sqrt{2}}\Big(\ket{v_\ell,H}_A\pm\ket{h_\ell,V}_A\Big),
\end{eqnarray}
where the first and second positions inside the ket represent the OAM and polarization states of photon \emph{A}, respectively, and $\ket{h_\ell}$ and $\ket{v_\ell}$ refer to the horizontal and vertical basis in the OAM subspace of $\{\ket{+\ell},\ket{-\ell}\}$. The Bell states are mutually orthogonal and form a complete basis in the spin-OAM Hilbert space of $\{\ket{+\ell,H},\ket{-\ell,H},\ket{+\ell,V},\ket{-\ell,V}\}$. Thus, we can rewrite Eq.~(\ref{SPDC_state}) in terms of the spin-OAM Bell states of Eq.~(\ref{bell_states}):
\begin{eqnarray}\label{state_bell}
 	\ket{\chi}&=&\frac{1}{2}\,\sum_{\ell=0}^{\infty}c_{\ell}
 	\bigg\{\ket{\Phi^+_\ell}_A\Big(\alpha\ket{h_\ell}_B+\beta\ket{v_\ell}_B\Big)\nonumber\\
	&+&\ket{\Phi^-_\ell}_A\Big(\alpha\ket{h_\ell}_B-\beta\ket{v_\ell}_B\Big)\nonumber\\
	&+&\ket{\Psi^+_\ell}_A\Big(\alpha\ket{v_\ell}_B+\beta\ket{h_\ell}_B\Big)\nonumber\\
	&+&\ket{\Psi^-_\ell}_A\Big(\alpha\ket{v_\ell}_B-\beta\ket{h_\ell}_B\Big)\bigg\}|H\rangle_B\,,
\end{eqnarray}
where we used $\ket{\pm\ell}=\left(\ket{h_\ell}\pm i \ket{v_\ell}\right)/{\sqrt{2}}$~\cite{padgett:99}. As can be seen from Eq.~(\ref{state_bell}), if \emph{A} performs one of the Bell-state measurements defined in Eq.~(\ref{bell_states}), the state of photon \emph{B} is left in a superposition of orthogonal OAM modes with the coefficients determined by the polarization state of photon \emph{A}. This means that depending on the polarization setting of photon \emph{A}, photon \emph{B} is in a different superposition of OAM modes. Assuming a fixed OAM basis, \emph{B} applies one of the unitary operators from the set: $\{\hat{\mathbb{1}},\hat{\sigma}_x,i\hat{\sigma}_y,\hat{\sigma}_z\}$ to photon \emph{B}'s OAM state based on the outcome of \emph{A}'s Bell-state measurement.
\begin{figure*}[ht]
 \center
  \includegraphics[width=0.9\textwidth]{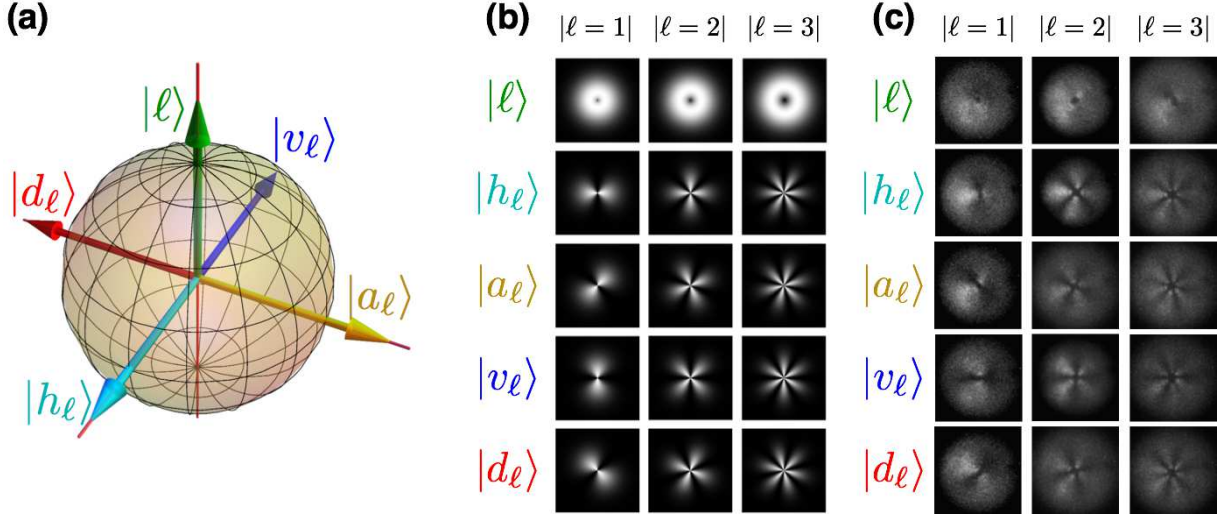}
	\caption{\label{fig:data1} (Online color) Qualitative comparison of experimental data and theoretical prediction. \textbf{(a)} Pictorial state representation of the teleported states on the OAM Poincar\'e sphere of $\{\ket{+\ell},\ket{-\ell}\}$. In this representation, south and north poles represent the $\ket{+\ell}$ and $\ket{-\ell}$-states, respectively, and an equal superposition of $\ket{+\ell}$ and $\ket{-\ell}$ stands on the equator. \textbf{(b)} Theoretically predicted spatial distributions of the teleported states corresponding to initial polarization states of \emph{C} being set to circular-left $\ket{L}\rightarrow\ket{\ell}$, horizontal $\ket{H}\rightarrow\ket{h_\ell}$, anti-diagonal $\ket{A}\rightarrow\ket{a_\ell}$, vertical $\ket{V}\rightarrow\ket{v_\ell}$, and diagonal $\ket{D}\rightarrow\ket{d_\ell}$. \textbf{(c)} Experimentally recorded spatial distributions of photon \emph{B} on the ICCD camera conditioned by detecting photon \emph{A} in the spin-orbit Bell state of $\ket{\Phi^+_\ell}$. Total exposure time per picture is $600$~s with a time window of $4$~ns.}
\end{figure*}

The projection of photon \emph{A} onto one of the four single-photon spin-OAM Bell states, i.e. $\ket{\Phi^{\pm}}$, $\ket{\Psi^{\pm}}$, can be achieved with a Sagnac based interferometer (shown as an inset in Fig.~\ref{fig:fig1}) followed by a half-wave plate and a spatial light modulator (SLM)~\cite{Slussarenko:10}. Indeed, the polarizing Sagnac interferometer in which a Dove prism is inserted (PSI-DP) couples spin-to-orbital angular momentum of the photon. Since the photon is collimated in the interferometer, the transformation of the Dove prism can be approximately described as acting only on the OAM space;  $\text{DP}_{\pm\theta}\cdot\ket{\pm\ell}\rightarrow e^{\pm 2 i\ell\theta}\ket{\mp\ell}$, where $\pm\theta$ is the rotational angle of the Dove prism with the sign depending on the propagation direction~\cite{Gonzalez:06, padget99}. The polarizing beam splitter (PBS) at the entry of the PSI-DP converts the incoming photon state into a superposition of two counter-propagating horizontally and vertically polarized states. Due to the presence of the rotated Dove prism these counter-propagating beams accumulate a relative phase difference of $|4\ell\theta|$. By setting $\theta=\pi/{\left(8\ell\right)}$, the PSI-DP transforms each of the four Bell states according to:
\begin{eqnarray}\label{eq:proj}
\left\{\begin{array}{c}
	\ket{\Phi^+} \rightarrow \ket{v_\ell,A}  \\ 
	\ket{\Phi^-} \rightarrow \ket{v_\ell,D}  \\
	\ket{\Psi^+} \rightarrow \ket{h_\ell,A}  \\
	\ket{\Psi^-} \rightarrow \ket{h_\ell,D}
\end{array}\right.,
\end{eqnarray}
where unnecessary global phases are omitted. It is worth mentioning that a recently invented liquid crystal device, the so-called $q$-plate, implemented with appropriate wave plates and a PBS can also be used to sort all spin-orbit Bell-states~\cite{marrucci:11}. In order to project the state of photon \emph{A} onto one of the Bell states, the outgoing photons from PSI-DP must be projected onto the transformed states given in Eq.~(\ref{eq:proj}). This can be achieved with a combination of a $\pi/8$-rotated HWP, an SLM, and a single mode optical fiber, to project onto the $h_\ell$ or $v_\ell$ states. The SLM displaying the desired hologram in conjunction with a single mode optical fiber selects a definite OAM sub-space $\abs{\ell}$. Indeed, this requires very precise aligning of both the near and far-field of two counter-propagating beams inside the PSI-DP in which the centers of the single mode optical fiber, the hologram and the SPDC source are precisely superimposed. In contrast to the case of multi-particle Bell states, the situation is different for single particle hybrid Bell states, where it is possible to perform a complete Bell-state measurement deterministically and with 100\% efficiency~\cite{calsamiglia2001maximum, kwiat1998embedded}. This can be accomplished with a suitable choice of settings for the HWP and SLM following the PSI-DP. In the present experiment, we choose to project onto the $\ket{\Phi^+_\ell}$ state because it does not require any additional operation by \emph{B}, and leaves \emph{B}'s photon in a superposition of OAM states described by Eq.~(\ref{state_bell}), i.e. $\ket{\chi}_{B}={}_A\!\braket{\Phi^+_\ell}{\chi}\propto\alpha\ket{h_\ell}_B+\beta\ket{v_\ell}_B$ . In other words, when $\ket{\Phi^+_\ell}$ is measured and the detector fires, \emph{B} finds his photon in the state $\alpha\ket{h_\ell}_B+\beta\ket{v_\ell}_B$, with the same coefficients $\alpha$ and $\beta$ as in the unknown polarization state.
\begin{figure*}[t]
	\center
	\includegraphics[width=0.95\textwidth]{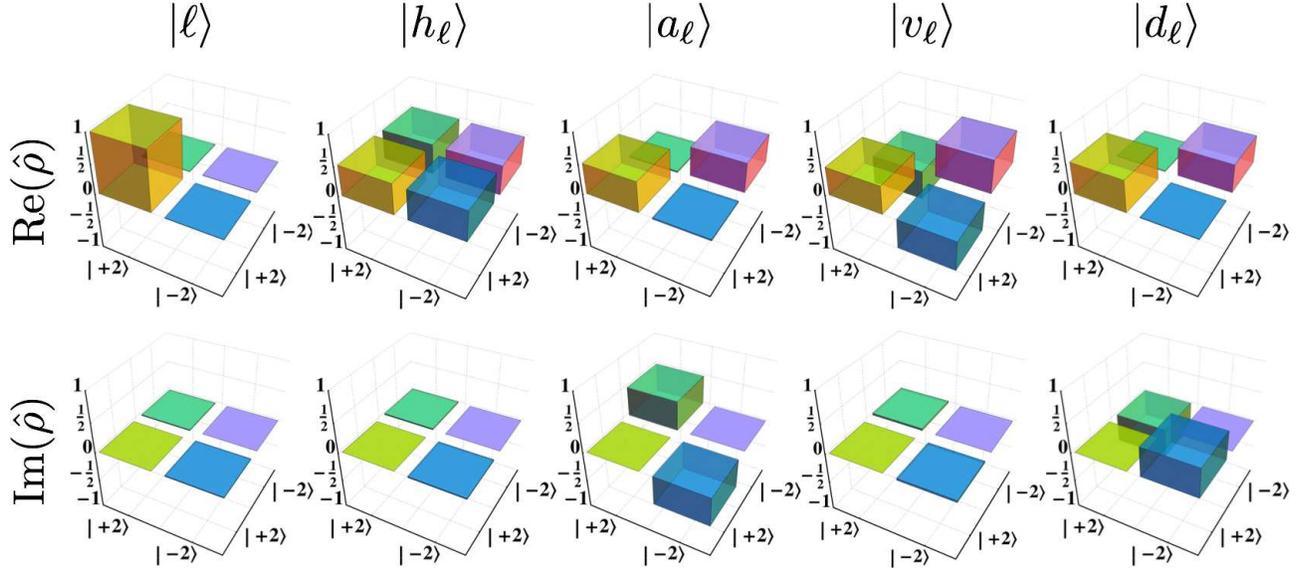}
	\caption{\label{fig:data2} (Online color) Real (upper row) and imaginary (lower row) parts of the reconstructed density matrices for the teleported states in OAM subspace of $\abs{\ell}=2$ shown in Fig.~\ref{fig:data1}. The density matrices are reconstructed via quantum state tomography, where projections over six eigenstates of Pauli matrices are used to estimate the four real parameters that specify the density matrix.}
\end{figure*}

In the paraxial approximation the $\ket{\pm\ell}$ states can be represented by Laguerre-Gauss modes, i.e. $\braket{\mathbf{r}}{\pm\ell}\propto\exp{(\pm i\ell\varphi)}$, where $\varphi$ is the azimuthal angle in polar coordinates. Due to the post-selection which is introduced by the single mode optical fiber, only the lowest radial $p$-index contributes; henceforth we only consider the $p=0$ radial mode. The theoretical prediction of the spatial distribution of \emph{B}'s photons on the ICCD camera is plotted in Fig.~\ref{fig:data1}-{\bf(b)}. By capturing \emph{B}'s photons with an ICCD camera, a qualitative comparison between the theoretical expectation and the experimental results can be made. To ensure that the entangled partner photon of \emph{A} is captured by the camera, an electronic signal is transmitted from \emph{A}'s detector to \emph{B}'s camera. To achieve this, photon \emph{A} is guided via a single mode optical fiber to a single photon detector, which generates a TTL-signal that triggers the ICCD camera (Andor iStar $1024\times1024$). Since the pump laser has a repetition rate of $100$ MHz, the maximum time window for capturing \emph{B}'s photon is $10$~ns. To further ensure that only  entangled partner photons of \emph{A}'s trigger photon are collected, we chose the detection time window of the ICCD camera to be $4$~ns.  The length of the delay line is chosen to compensate for the electronic delay which is introduced by the detector and ICCD camera. In our case this is approximately $100$~ns. ICCD camera has recently been used to visualize several fundamental quantum experiments such as the quantum entanglement~\cite{fickler:13}, ghost imaging~\cite{aspden:13} and Popper's thought~\cite{eliot:14} experiment. While photon \emph{B} is circulating in the delay line, \emph{C} prepares the state which he is going to teleport to \emph{B}. With the sequence of QWPs and HWPs, the coefficients $\alpha$ and $\beta$ are set to $\alpha=\sin{(\gamma/2)}$ and $\beta=\cos{(\gamma/2)}\,e^{i\delta}$, where $\gamma$ and $\delta$ are azimuth and polar angles of the polarization state on the Poincar\'e sphere~\cite{karimi:10b}. 

In order to examine our teleportation scheme qualitatively, we can compare the theoretically predicted probability distribution of teleported states with the experimental data taken using the ICCD camera, with the wave plates set at different angles. Figure~\ref{fig:data1}-{\bf(c)} shows the \emph{joint} probability distribution of teleported states of photon \emph{B}. All images are captured in the far field of the SPDC source, and a so-called sector hologram for projecting onto the $\ket{v_\ell}$ is displayed on the SLM. The sector hologram is generated by imprinting the corresponding phase distribution of $\ket{v_\ell}=-i\left(\ket{+\ell}-\ket{-\ell}\right)/\sqrt{2}\propto\sin{(\ell\varphi)}$ state onto a normal blazed grating, i.e. $\text{Mod}\left(\text{sgn}\left({\sin{(\ell\varphi)}}\right)+2\pi x/\Lambda,2\pi\right)$ where Mod is the modulo function that gives the reminder of the first argument divided by the second one, sgn is the sign function, $\Lambda$ and $x$ are the grating pitch and the cartesian coordinate, respectively. As can be seen, apart from contrast quality, the theoretical predictions Fig.~\ref{fig:data1}-{\bf(b)} and the experimental data Fig.~\ref{fig:data1}-{\bf(c)} are in good agreement. We can also rule out superpositions of higher order $p$-modes, because the single mode optical fiber used for triggering the ICCD camera filters the photons from higher order $p$-modes. 

The images taken with the ICCD camera shown in Fig.~\ref{fig:data1} qualitatively illustrate the results of this hybrid teleportation scheme, and are not sufficient to determine the quality of the teleported states. For the sake of completeness, we also measure the fidelity to estimate the quality of the teleportation scheme. This can be done by performing quantum state tomography on the teleported states.  We utilized projections onto states from mutually unbiased bases (MUBs) for a bi-dimensional Hilbert space $\{\ket{+\ell},\ket{-\ell} \}$. The measurements consist of projections onto states from the set $\{h_\ell, v_\ell, a_\ell, d_\ell, l_\ell, r_\ell\}$, where the index $\ell$ represents the OAM working subspace~\cite{jack2009precise}. These states are eigenstates of the Pauli matrices. With those measurements, the density matrix of the state can be reconstructed using the maximum likelihood estimate. In order to perform these projective measurements conditioned by \qo{clicks} on the detector \emph{A}, we replaced the ICCD camera with a second SLM followed by a single mode optical fiber and a second APD (there is no need for keeping the photon in the delay line). The coincidence counts between the two detectors are measured by means of a coincidence box with a time window of $10$~ns.  The second SLM on arm \emph{B} projects the photons onto one of the aforementioned states. The density matrices corresponding to the different teleported states are reconstructed via this \emph{over-complete} set of measurements. All coincidence counts for reconstructing the density matrices are averaged over $100$~seconds. 

\begin{table}[t]
  \centering
  \caption{\label{tab:tab1} Measured fidelity $F$ of different teleported states in the OAM subspace of $\abs{\ell}=2$. The last two states, i.e. $\zeta$ and $\eta$, are arbitrary elliptical polarization states obtained by rotation of a single quarter-wave plate.\vspace{2mm}}
  \begin{tabular*}{0.46\textwidth}{@{\extracolsep{\fill}} c|c|c}
  \hline\hline
 Initial polarization state & Teleported state to Bob & $F$  \\ \hline\hline
 $\ket{L}$ & $\ket{+2}$ & $0.995\pm0.003$ \\\hline
 $\ket{V}$ & $\ket{v_2}$ & $0.994\pm0.004$ \\\hline
 $\ket{D}$  & $\ket{d_2}$  & $0.984\pm0.008$ \\\hline
 $\ket{H}$  & $\ket{h_2}$ & $0.999\pm0.002$ \\\hline
 $\ket{A}$  & $\ket{a_2}$ & $0.992\pm0.013$ \\\hline
 $\ket{R}$  & $\ket{-2}$ & $0.999\pm0.001$ \\\hline
 $\ket{\zeta}$  & $\ket{\zeta_2}$ & $0.997\pm0.005$\\\hline
 $\ket{\eta}$  & $\ket{\eta_2}$ & $0.999\pm0.005$\\
 \hline\hline
 \end{tabular*}
\end{table}

The density matrices for the different teleported states are shown in Fig.~\ref{fig:data2}. The key performance indicator of a successful teleportation is the state fidelity, which is defined as $F= \left(\hbox{Tr}\sqrt{\sqrt{\hat{\rho}}\;\hat{\rho}_r\sqrt{\hat{\rho}}}\;\right)^2$, where $\hat{\rho}_r$ and $\hat{\rho}$ are the reconstructed and theoretical density matrices, respectively. Fidelity of different teleported states is reported in Tab.~\ref{tab:tab1}. All fidelities are above 98.4\% which means that the quality of the teleported states is very high, and confirms the validation of our examined scheme.  

In summary, we have experimentally shown that it is possible to teleport a generally unknown quantum state from spin angular momentum space of a single photon onto an OAM subspace. A significant advantage of this teleportation scheme is that only two particles are involved, so there is no need for a third ancilla particle. Instead of an ancilla particle, a different degree of freedom of one of the entangled particles was used. Moreover, unlike the multi-particles teleportation scheme, our proposal is deterministic. The very high fidelities of the teleportation scheme and the relatively simple experimental technique required with an efficiency approaching 100 \% make this specific teleportation scheme very promising for quantum key distribution, and in general for quantum cryptography. Our work also opens the possibility to use orbital angular momentum of photons in different quantum computing applications. Either as qudits in the computing process itself or as a connecting system between optical quantum computers, driven by photons with polarization encoded qubits, and photonic quantum memories.

The authors thank Filippo Cardano for fruitful discussion, and acknowledge the support of the Canada Excellence Research Chairs (CERC) Program. R. W. B. also acknowledges the support by DARPA.

\end{document}